\documentclass[11pt]{article}
\usepackage{amsmath,amssymb,color,epsf}
\usepackage{amssymb,slashed,latexsym,amsmath,multirow,color,dsfont}
\usepackage{graphics}
\usepackage{graphicx}
\usepackage{cite}

\makeatletter
\@addtoreset{equation}{section}
\makeatother

\usepackage{amsfonts}

\usepackage{float}

\usepackage{soul}

\usepackage{amsfonts}

\usepackage[font=footnotesize,labelsep=newline,labelfont=sc,justification=centering,position=top]{caption}

\newcommand{\bea}{\setlength\arraycolsep{2pt} \begin{eqnarray}}
\newcommand{\eea}{\end{eqnarray}}
\newcommand{\nn}{\nonumber}

\usepackage{hyperref}

\newsavebox{\uuunit}
\sbox{\uuunit}
    {\setlength{\unitlength}{0.825em}
     \begin{picture}(0.6,0.7)
        \thinlines
        \put(0,0){\line(1,0){0.5}}
        \put(0.15,0){\line(0,1){0.7}}
        \put(0.35,0){\line(0,1){0.8}}
       \multiput(0.3,0.8)(-0.04,-0.02){12}{\rule{0.5pt}{0.5pt}}
     \end {picture}}

\def\be{\begin{equation}}
\def\ee{\end{equation}}
\def\ba{\begin{array}}
\def\ea{\end{array}}
\def\bea{\begin{eqnarray}}
\def\eea{\end{eqnarray}}
\def\bd{\begin{displaymath}}
\def\ed{\end{displaymath}}

\def\nn{\nonumber}

\def\a{\alpha}
\def\b{\beta}
\def\g{\gamma}

\def\d{\delta}

\def\l{\lambda}

\def\m{\mu}
\def\n{\nu}
\def\r{\rho}

\def\o{\omega}

\def\nn{\nonumber}

\def\cL{\mathcal{L}}

\DeclareMathOperator{\Tr}{Tr}

\begin{document}

\begin{titlepage}

\begin{center}

\hfill UG-16-73 \\  

\vskip 1.5cm

{\Large \bf Holographic Entanglement Entropy in NMG}
\vskip 1cm

{\bf Luca Basanisi\,$^1$, Shankhadeep Chakrabortty\,$^1$. } \\

\vskip 25pt

{\em $^1$ \hskip -.1truecm Van Swinderen Institute for Particle Physics and Gravity,  \\ 
University of Groningen, Nijenborgh 4, 9747 AG Groningen, The Netherlands \vskip 5pt }

{email: {\tt l.basanisi@rug.nl, s.chakrabortty@rug.nl} } \\

\vskip 25pt

\end{center}

\vskip 0.5cm

\begin{center} {\bf ABSTRACT}\\[3ex]
\end{center}

In this paper, we show that a higher derivative theory, such as New Massive Gravity, allows the existence of new entangling surfaces with non-zero extrinsic curvature. We perform the analysis for Lifshitz and Warped~$AdS$ spacetimes, revealing the role of the higher derivative contributions in the calculation of the holographic entanglement entropy.  Finally, as an outcome of our holographic analysis we briefly comment on the dual boundary theory.

\end{titlepage}

\tableofcontents

\setcounter{page}{1}

\section{Introduction}

Entanglement entropy (EE) is conceived as a very general tool to measure the quantum correlation between two systems. It encodes the amount of information loss when one of the two systems becomes inaccessible. This non-local measure plays a crucial role as an order parameter to probe the quantum phase transition in many physical contexts~\cite{Calabrese:2004eu}. To elaborate the idea of entanglement entropy, let us consider a bipartite system described by a well-defined Hilbert Space~$\mathcal{H}_{tot}$  such that this can be 
factorized into two disjoint Hilbert spaces of the subsystems A and B as, 
\begin{eqnarray}
\mathcal{H}_{tot} = \mathcal{H}_A \otimes \mathcal{H}_B.
\end{eqnarray}
For the observer who has the access of only the region~$A$, the system is effectively represented by  the reduced density matrix~$\rho_A$,
\begin{eqnarray}
\rho_{A} = \Tr_{B}\, \rho_{tot},
\end{eqnarray}
where the partial trace is performed only over~$\mathcal{H}_B$. Here, $\rho_{tot}$ is the density matrix characterizing the full system .
The entanglement entropy of the subsystem~$A$ is defined after the von Neumann entropy of the reduced density matrix~$\rho_{A} $,
\begin{eqnarray}\label{eq:EElog}
S_A = - \Tr_{A}\, \rho_{A}\, \log \rho_{A}.
\end{eqnarray}

Despite the simplicity of this formula and its successful application to simple quantum mechanical systems, it is extremely difficult to generalize the prescription~(\ref{eq:EElog}) to the perturbative quantum field theories in arbitrary dimensions. However, for two dimensional CFT, the symmetry structure of the theory encourages us to apply the replica trick~\cite{Calabrese:2004eu} to overcome the technical problems one might encounter. Here, we can compute the R$\acute{\text{e}}$nyi entropy~($S_n = \frac{1}{n-1} \log \Tr \rho^n_A$) of~$n$ copies of the system and then take the limit~$n\rightarrow1$ to obtain the EE. However, for higher dimensional conformal field theories, the replica method can be applicable only for certain topologies of the entangling region. It is also important to note that, although the presence of infinitely many degrees of freedom in field theory makes this quantity divergent, it can be regularized by introducing a UV cut-off~\cite{Calabrese:2004eu, Casini:2009sr}.   

The analysis of entanglement entropy in strongly coupled quantum systems requires techniques beyond the perturbative regime. However, these obstacles can be overcome by considering a holographic realization of Entanglement Entropy (HEE) originally proposed by Ryu and Takayanagi (RT) in their seminal work~\cite{Ryu:2006bv, Ryu:2006ef}. According to their proposal, the entanglement entropy~$S_A$ of a region~$A$ in a~$d$ dimensional boundary theory corresponds holographically to a geometrical quantity, i.e. the area of a codimension-$2$ spacelike minimal surface~$\gamma_A$ in the~$d+1$ dimensional dual gravity theory. The minimal surface is anchored to the boundary in such a way that it satisfies the homology constraint~$\partial \gamma_A= \partial A$. The exact statement of their proposal is astonishingly simple and reads as 
\begin{equation}\label{EE}
S_A = \frac{{\rm Area} (\g_A)}{4 G_N^{(d+1)}}\,,
\end{equation}
where~$G_N^{(d+1)}$ is the $d+1$ dimensional Newton constant. The generalization of this formula for asymptotically~$AdS$ static spacetimes has been achieved~\cite{Nishioka:2009un}. Furthermore, the covariant version of the RT proposal for time dependent background has been formulated in~\cite{Hubeny:2007xt}.

Simplicity is not the only wonderful feature of this proposal and not even the most interesting one. Indeed, one can interpret the equation~(\ref{EE}) as an indication that the quantum properties of matter are deeply related to a geometrical object on the other side of the correspondence. Such indication yields a fascinating perspective in the context of emergent spacetimes~\cite{VanRaamsdonk:2010pw,Swingle:2009bg}.

Recently, a holographic proof of the RT proposal has been expounded by Lewkowycz and Maldacena (LM) in~\cite{Lewkowycz:2013nqa}. The essence of the proof is established by implementing the~$n$-copy replica trick in the dual bulk geometry. The metric of this replicated bulk geometry acquires a~$Z_n$ singularity on the hypersurface. The powerfulness of this method reveals that, by imposing the limit~$n\rightarrow1$, the hypersurface converges to the usual minimal surface in the RT proposal. 

It is then a very appropriate question to ask whether the LM formalism remains valid beyond the Einstein-Hilbert theory.  The effective description of the UV limit of the Einstein-Hilbert theory comprises higher derivative terms and turns out to be one of the most natural arenas for this investigation.  The LM formalism instigates the generalized formulations of holographic entanglement entropy for various higher derivative theories~\cite{Dong:2013qoa, Fursaev:2013fta, Camps:2013zua}. These generalizations allow contributions coming from the Wald's entropy~\cite{Wald:1993nt} as well as from the extrinsic curvature evaluated on the entangling surface. In particular, for our present analysis we follow the prescription of~\cite{Dong:2013qoa} for the holographic computation.

Among many successful formulations of higher derivative theories in various dimensions, three dimensional New Massive Gravity (NMG)~\cite{Bergshoeff:2009hq} draws substantial attention due to its wide class of background solutions. In addition to that, since such a theory lives in only three dimensions, the co-dimension~$2$ surface is just a line and thus the technical obstacles to carry out the actual holographic computations are drastically reduced. However, due to the presence of higher derivative terms, NMG provides a structure complex enough to observe non trivial changes in the behaviour of EE, since we will show the existence of new minimal surfaces.

In the context of NMG, it has been pointed out in~\cite{Erdmenger:2014tba} that to specify the appropriate entangling surface by extremizing the entropy functional, a supplementary condition is essential for physical consistency in addition to usual homology constraint.  However, to find out a unique entangling surface anchored to the boundary by extremizing the higher derivative functional given in~\cite{Dong:2013qoa} remains unresolved to the authors due to the occurrence of insufficient number of boundary conditions. For~$AdS_3$ background in NMG, it turns out that the respective geodesic satisfies the higher derivative equation of motion. However, for a specific range of NMG parameter, in~\cite{Ghodsi:2015gna} it is shown that a new entangling surface, giving a lower entanglement entropy, exists. Moreover, in~\cite{Alishahiha:2013dca} the authors have shown that, for a particular class of asymptotically~$AdS$ spacetimes in four dimensional higher derivative theory, a perturbative modification of~$AdS$ geodesic provides the correct entangling surface satisfying  the necessary physical boundary conditions. Further analysis in this context can be found in~\cite{Bhattacharyya:2013gra,Hosseini:2015vya,Hosseini:2015gua}.

Motivated along this line of research, we compute holographic entanglement entropy for Lifshitz and Warped~$AdS$ backgrounds in New Massive Gravity.  Similar to~\cite{Alishahiha:2013dca}, we adopt the perturbative approach for our analysis. We observe that a suitable perturbative ansatz for the entangling surface significantly reduces the technical complexities of extremizing the higher derivative entropy functional. In particular, we study the non-trivial modification of the~$AdS_3$ geodesic by introducing a suitable perturbation resulting from the higher derivative contribution in the NMG theory. The modified entangling surface satisfies the equation of motion derived from the entropy functional order by order with a certain set of appropriate boundary conditions. We also give an interpretation of our holographic analysis consistent with the corresponding boundary theory.

The paper is organized as follows. In section~\ref{section:theory}, we briefly introduce the necessary ingredients to determine the geometry of the entangling surface and thus compute the entanglement entropy. In section~\ref{section:AdS}, we reproduce the analysis presented in~\cite{Ghodsi:2015gna} in order to review the procedure in the simple example provided by the Anti de-Sitter spacetime. We then apply the same technique to the Lifshitz spacetime in section~\ref{section:Lif}. In this background, we will prove the existence of a new entangling surface by deforming the geodesic. Consequently, we establish that the existent analysis in this regard~\cite{Hosseini:2015gua} requires further attentions. As a final example, in section~\ref{section:WAdS} we investigate the deformation of the entangling surface for the Warped~$AdS_3$ spacetime. Previous works on this particular case can be found in~\cite{Anninos:2013nja,Castro:2015csg,Ghodrati:2016ggy}, but our analysis will focus specifically on the higher derivative contribution. Finally, we summarize and discuss our results in section~\ref{section:disclusion}.

\section{NMG and Holographic Entanglement Entropy} \label{section:theory}

New Massive Gravity~\cite{Bergshoeff:2009hq} is a parity-preserving theory describing gravity in three dimensions that allows a massive graviton. Such a modification of the pure gravity theory is realized by adding a combination of higher derivative terms. In particular, the action, in the Euclidean signature, takes the following form
\begin{equation}\label{eq:NMG}
S = - \frac{1}{16 \pi G} \int d^3 x \sqrt{g} \left[ R + \frac{2}{L^2} + \frac{1}{m^2} \left(R_{\m\n} R^{\m\n} - \frac{3}{8} R^2 \right) \right]\,.
\end{equation}
One of the merits of this theory is that, although it lives in three dimensions, it presents a richer dynamics compared to Einstein gravity without introducing many of the well-known pathologies that we can find in a general four dimensional higher derivative theories.

Since New Massive Gravity is a higher derivative theory with curvature squared terms, we need a reassessment of the Ryu-Takayanagi prescription. It is known that the finite part of the holographic entanglement entropy evaluated in a black hole background in the Einstein gravity corresponds to the Bekenstein-Hawking thermal entropy~\cite{Nishioka:2009un, Fischler:2012ca}. Therefore it is natural to expect the same in gravity theories with higher derivatives, with the understanding that the thermal entropy is now realized as Wald's entropy~\cite{Wald:1993nt}. Recently, motivated by the analysis in~\cite{Lewkowycz:2013nqa}, a general prescription for computing the holographic entanglement entropy for higher derivative theory was proposed in~\cite{Dong:2013qoa}. When applied to New Massive Gravity, the prescription~\cite{Dong:2013qoa} yields the entropy functional~\cite{Bhattacharyya:2013gra}
\begin{equation} \label{eq:EEfunc}
S_{EE} = \frac{1}{4G} \int_{\Sigma} dz\, \sqrt{h} \, \left[ 1 + \frac{1}{m^2} \left(R_{||} - \frac{1}{2} K^2 - \frac{3}{4} R \right) \right]\,,
\end{equation}
where~$h$ is the induced metric on the entangling surface~$\Sigma$. Such surface is taken to be co-dimension two (thus it is just a line), anchored to the boundary and propagating deep in the bulk.  The projected Ricci tensor $R_{||}$ is given by 
\begin{equation}
R_{||} = \eta^{\a\b} (n_{(\a)})^{\m} (n_{(\b)})^{\n} R_{\m \n}\,,
\end{equation}
while~$(n_{(\a)})^{\m}$ are the orthogonal vectors defined on~$\Sigma$. The extrinsic curvature is given by
\begin{equation}
(K_{(\a)})_{\m\n} = h_{\m}^{\l} h_{\n}^{\rho} \nabla_{\rho} (n_{(\a)})_{\l}\,,
\end{equation}
where~$\nabla$ is the covariant derivative with respect to the bulk metric.  Consequently, the contracted form of the extrinsic curvature entering the functional can be written as
\begin{equation}
K^2 = \eta^{\a\b} (K_{(\a)})_{\m}^{\m} (K_{(\b)})_{\n}^{\n}\,.
\end{equation}

All these quantities are required to be evaluated on the appropriate entangling surface, determined by minimizing the entropy functional~(\ref{eq:EEfunc}) itself. In three dimensional pure gravity this problem is easily solved by taking the geodesic as entangling surface since it is, by definition, the curve with minimal length.  However, as we will see in the next sections, this is not necessarily the case for a more general theory of gravity. Intuitively, due to the presence of higher derivative terms, the entropy functional given in~(\ref{eq:EEfunc}) fails to be interpreted as a length anymore~\cite{Erdmenger:2014tba}. On the technical point of view, minimizing such functionals leads to a higher order differential equation that opens up the possibility of finding different entangling surfaces as opposed to the one in the context of Einstein-Hilbert gravity. In the next section, we elaborate upon this issue further with a concrete example, i.e. the~$AdS_3$ spacetime as a background in the New Massive Gravity.   
\newpage

\section{New Entangling Surfaces} \label{section:AdS}

In this section, we elaborate on the geometry of the entangling surface embedded in a three dimensional background describing a more general theory of gravity, in particular New Massive Gravity. Our principal aim is to achieve an entangling surface by minimizing the entropy functional prescribed in~\cite{Dong:2013qoa} and to compute the corresponding holographic entanglement entropy. The simplest example to realize the richer behaviour of the geometry of the entangling surface we are interested in is the~$AdS_3$ spacetime. Such background is described by the metric
\begin{equation}
 d s^2 = g_{\m\n} d x^{\m} dx^{\n} = \frac{\widetilde{L}^2}{z^2} (dt^2 + dz^2 + dx^2)\,,
\end{equation}
where~$\widetilde{L}$ is the~$AdS_3$ radius and it is related to the cosmological parameter~$L$ present in~(\ref{eq:NMG}) by
\begin{equation} \label{eq:Fdef}
L^2 = F \widetilde{L}^2\,, \qquad \qquad F^2  - 4 m^2 L^2 F + 4 m^2 L^2 = 0\,.
\end{equation}
With this background metric, we systematically reproduce the results presented in~\cite{Ghodsi:2015gna} in order to give a clear overview of the general procedure. The entangling region in the dual field theory is a one dimensional line located at the boundary of the~$AdS_3$~($z=0$).  Correspondingly, to obtain the co-dimension~$2$ extremal hypersurface embedded in the constant time slice of~$AdS_3$ geometry, we choose the following ansatz consistent with the so-called boundary parametrization
\begin{equation}
t = 0 \,, \qquad \qquad x = f(z)\,.
\end{equation}

With this~$AdS_3$ background metric and the prescribed profile ansatz for the entangling surface, the computation of the entropy functional~(\ref{eq:EEfunc}) (we refer to~\cite{Ghodsi:2015gna} for the details of the calculation) leads to
\begin{equation}\label{eq:AdSEE}
 S_{EE} = \frac{2 \pi}{\ell_p} \int dz \, \frac{\widetilde{L}}{z} \, \sqrt{A} \left[ 1 + 2 \frac{F-1}{F} \left[1 - \frac{1}{A^3} \left(f'(z)^3 + f'(z) - z f''(z) \right)^2 \right] \right]\,,
\end{equation}
where~$A = f'(z)^2 + 1$. In order to minimize this functional, we consider~(\ref{eq:AdSEE}) as a one dimensional action, therefore the corresponding equation of motion is
\begin{equation}\label{eq:adseom}
 \frac{\partial^2}{\partial z^2} \left(\frac{\d \cL}{\d f''}\right) - \frac{\partial}{\partial z} \left(\frac{\d \cL}{\d f'}\right) + \frac{\d \cL}{\d f} = 0\,.
\end{equation}
Since~(\ref{eq:AdSEE}) is independent of~$f(z)$, the last term in~(\ref{eq:adseom}) identically vanishes. The resultant fourth order differential equation is of highly non-linear nature. However, as mentioned  in~\cite{Bhattacharyya:2013gra}, there exists a very simple analytic solution of~(\ref{eq:adseom}), namely  
\begin{equation} \label{eq:geo}
 f_1(z) = \sqrt{z_0^2 - z^2}\,.
\end{equation}
where~$z_0=f(z=0)$ is a tunable parameter, expressing the length of the region of our interest. The turning point, that is how deep the surface goes into the bulk, is located at~$z_t=z_0$. Moreover, the corresponding extrinsic curvature vanishes. It is very interesting to note that the same solution can be obtained even without the higher derivative terms, being a geodesic anchored to the boundary region of our interest. As mentioned before, the profile~(\ref{eq:geo}) is the only possible solution in Einstein gravity, because it is the trajectory that minimizes the length. However, there is no a priori reason to expect that no solution other than~(\ref{eq:geo}) exists for a higher derivative theory of gravity.

Indeed, the authors of~\cite{Ghodsi:2015gna} present a different entangling surface. Making the ansatz~$f(z) = \sqrt{z_0^2 - z^2 + a z}$, and requiring that it solves the differential equation~(\ref{eq:adseom}), we obtain
\begin{equation}\label{eq:sol2}
 f_2(z) = \sqrt{z_0^2 - z^2 + 2 z_0 q z}\,, \qquad \qquad q = \sqrt{\frac{F-2}{F}}\,.
\end{equation}
In this case, the turning point goes deeper into the bulk and it is located at
\begin{equation}
z_t = z_0 \left[ q + \sqrt{q^2 + 1} \right]\,.
\end{equation}
Moreover, unlike the case for~$f_1(z)$, the contracted form of the extrinsic curvature evaluated on this new extremal surface 
\begin{equation}
 K^2 |_{f_2 (z)} = \frac{q^2}{ \widetilde{L}^2\, (q^2 + 1 )  }\, = \, \frac{F-2}{2 \widetilde{L}^2\,(F-1)}\,,
\end{equation}
is non-vanishing.

For both solutions we have the respective universal terms
\begin{eqnarray}
 S_{EE}^{(1)} = \frac{c_1}{3} \log \left(\frac{z_0}{\epsilon} \right)\,, &\qquad&  \frac{c_1}{3} = \frac{L}{4G} \, \frac{3F - 2}{F^{\frac{3}{2}}}\,, \nn \\
 S_{EE}^{(2)} = \frac{c_2}{3} \log \left(\frac{z_0}{\epsilon} \right)\,, &\qquad&  \frac{c_2}{3} = \frac{L}{4G} \, \sqrt{8\frac{F - 1}{F^2}}\,.
\end{eqnarray}
It is easy to verify that as ~$F>2$, the coefficients~$c_1$ and~$c_2$ follow the inequality~$c_1 \geq c_2 > 0$. Therefore, in contrast with the results previously presented in the literature, the solution that minimizes the entropy is~$f_2 (z)$. Both solutions coincide for~$F=2$, as can be verified from the equation~(\ref{eq:sol2}). There exists another range of parameters ($2 \geq F \geq \frac{2}{3}$) where only the first solution is real valued and correspondingly~$c_1$ is a positive number. It is important to notice that~$c_1$ can be interpreted as the central charge from the boundary theory prospective~\cite{Bhattacharyya:2013gra}. Therefore, the boundary EE is expected to be reproduced by the holographic computation performed by taking~$f_1(z)$ as entangling surface. Moreover, in~\cite{Mozaffar:2016hmg} the authors used the field-redefinition invariance to restrict the admissable entangling surfaces to those with vanishing trace of extrinsic curvature, thus giving an argument in favor of the first type of solution.\footnote{We thank the authors of~\cite{Mozaffar:2016hmg} for the clarifying discussion on this aspect.}

We dedicate the next sections to investigate what happens in the more complex backgrounds that NMG provides us.

\section{HEE for Lifshitz spacetime in NMG} \label{section:Lif}

The next solution of New Massive Gravity that we want to consider is the Lifshitz background. The isometry of the  Lifshitz  spacetime can be holographically mapped to the symmetry of the dual non-Lorentz invariant boundary theories~\cite{Hartong:2014oma,Griffin:2012qx}. It thus offers a substantial understanding of strongly coupled nonrelativistic conformal field theories characterizing a large class of condensed matter systems~\cite{Son:2008ye,Balasubramanian:2008dm,Kachru:2008yh}. Moreover, for the purposes of our analysis, this case presents a lot of technical similarities to the AdS spacetime. Therefore, it seems that the Lifshitz spacetime is the natural choice to show the existence of new entangling surfaces in a more general context.

It is important to notice that, being an asymptotically non-$AdS$ background, the Lifshitz spacetime needs special attentions for holographic computations. In the context of Einstein-Hilbert theory coupled to matter field, the authors of~\cite{Korovin:2013bua} have constructed the bulk-to-boundary dictionary for the Lifshitz spacetime by treating it as a deformation over $AdS$. In particular, the authors have considered a perturbative expansion with respect to the Lifshitz exponent around unity. In this scenario, the dual boundary theory is a deformed conformal field theory consistent with the Lifshitz symmetry. In the following analysis, with the similar spirit of~\cite{Korovin:2013bua}, we perform the bulk analysis of the entanglement entropy for the Lifshitz spacetime in the context of NMG.  Another example of obtaining the holographic entanglement entropy for Lifshitz spacetime in the Lovelock gravity can be found in~\cite{deBoer:2011wk}.

The metric of the Lifshitz background is given by
\begin{equation}
 d s^2 = g_{\m\n} d x^{\m} dx^{\n} = \frac{\widetilde{L}^{2\nu}}{z^{2\nu}} dt^2 + \frac{\widetilde{L}^{2}}{z^{2}} (dz^2 + dx^2)\,,
\end{equation}
where~$\widetilde{L}$ is the Lifshitz radius and~$\nu$ is the Lifshitz exponent. Note that the limit~$\nu \rightarrow 1$ leads to the AdS spacetime discussed in the previous section. As well explained in~\cite{AyonBeato:2009nh}, the exponent~$\nu$ and the NMG parameters are related by
\begin{equation}
 m^2 \widetilde{L}^2 = \frac{1}{2} \left( \nu^2 - 3 \nu + 1 \right)\,, \qquad \qquad \frac{\widetilde{L}^2}{L^2} = \frac{1}{2} \left( \nu^2 + \nu + 1 \right)\,.
\end{equation}
The extremal surface, parametrized by the following relations
\begin{equation}
 t = 0 \,, \qquad \qquad x = f(z)\,,
\end{equation}
leads to the induced metric
\begin{equation}
ds_{h}^2 = h_{\m\n} d x^{\m} dx^{\n} = \frac{\widetilde{L}^2}{z^2} \left( f'(z)^2 + 1 \right) dz^2\,.
\end{equation}
We note that the induced metric in the present context is structurally identical to the one we find in the~$AdS_3$ spacetime. Although the timelike orthogonal vectors, defined on the co-dimensional two entangling surface, posses an explicit dependence of the Lifshitz exponent~$\nu$
\begin{equation}
 n_{\a\,1} = \left(0, -\frac{\widetilde{L} f'(z)}{z \sqrt{A}}, \frac{\widetilde{L}}{z \sqrt{A}}\right)\,, \qquad n_{\a\,2} = \left(\frac{\widetilde{L}^{\nu}}{z^{\nu}}, 0, 0 \right)\,, \qquad A = f'(z)^2 + 1\,,
\end{equation}
the components of the extrinsic curvature are the same as before. As the metric is diagonal, we are only interested in the diagonal terms of the extrinsic curvature. Since we have a Killing vector in the time direction, the component of the extrinsic curvature in that direction,~$K^2_{\a\a}$, vanishes. On the other hand, once we compute~$K^1_{\a\a}$, it is easy to verify that~$K^1_{tt} = 0$ and~$K^1_{rr} = f'(z)^2 K^1_{zz}$. So the component we need to know is
\begin{equation}
 K^1_{zz} = \frac{\widetilde{L}}{z^2 A^{5/2}} \left[ f'(z)^3 + f'(z) - z f''(z) \right]\,,
\end{equation}
leading to
\begin{equation}
 K^2 = \frac{1}{\widetilde{L} A^3} \left[ f'(z)^3 + f'(z) - z f''(z) \right]^2\,.
\end{equation}
Since the Lifshitz and the Anti de-Sitter spacetimes differ only in the~$g_{tt}$ component and we are working on a time slice, there is no difference in the induced metric and in the extrinsic curvature of the two cases. However the intrinsic curvature is a quantity that does not depend on the embedding, therefore it presents differences with respect to the~$AdS_3$ case, namely
\begin{equation}
 R = -\frac{2\left( \nu^2 + \nu + 1 \right)}{\widetilde{L}^2}\,, \qquad \qquad R_{||} = - \left[ \frac{\nu \left( \nu f'(z)^2 + 1 \right)}{\widetilde{L}^2\, A}\, + \frac{\left( \nu^2 + \nu + 1 \right)}{\widetilde{L}^2} \right]\,.
\end{equation}

Collecting all these results together and plugging them back into~(\ref{eq:EEfunc}), we obtain the entropy functional for the Lifshitz spacetime
\begin{eqnarray}\label{eq:eeLif}
 S_{EE} = \frac{1}{4G} \int dz \frac{\widetilde{L}}{z} \sqrt{A} \biggr[ 1 + \frac{2}{\nu^2 - 3 \nu + 1} \biggr[ &-& \frac{\nu \left( \nu f'(z)^2 + 1 \right)}{A} + \frac{\left( \nu^2 + \nu + 1 \right)}{2}  \nn \\
 &-& \left.\left. \frac{1}{2A} \left(f'(z)^3 + f'(z) - z f''(z) \right)^2 \right] \right]\,.
\end{eqnarray}
Upon minimizing the functional~(\ref{eq:eeLif}) in the same way as described in section~\ref{section:AdS}, it leads again to a highly non-linear differential equation
\begin{eqnarray}\label{eq:diffLif}
 &&(2 \nu -1) f'(z)^9+(2 \nu  (\nu +3)-3) f'(z)^7+(6 \nu  (\nu +1)-3) f'(z)^5 \nn \\
 &&+\left(4 \nu ^2-6 \nu +3\right) z f'(z)^6 f''(z)+z \left(-5 z^2   f''(z)^3-2 \nu ^2 f''(z)+2 z \left(z f^{(4)}(z)+2 f^{(3)}(z)\right)\right) \nn \\
 &&+f'(z)^3 \left(-5 z^2 f''(z) \left(4 z f^{(3)}(z)+3 f''(z)\right)+6 \nu ^2+2 \nu -1\right) \nn \\
 &&+f'(z) \left(2 \nu ^2-5 z^2 f''(z) \left(4 z f^{(3)}(z)+3 f''(z)\right)\right) \nn \\
 &&+ z f'(z)^2 \left(30 z^2 f''(z)^3+(3-6 \nu )f''(z)+4 z \left(z f^{(4)}(z)+2 f^{(3)}(z)\right)\right) \nn \\
 &&+2 z f'(z)^4 \left(3 (\nu -1)^2 f''(z)+z \left(z f^{(4)}(z)+2 f^{(3)}(z)\right)\right) = 0\,.
\end{eqnarray}

In~\cite{Hosseini:2015gua}, it is stated that this equation is solved by the geodesic, i.e.
\begin{equation} \label{eq:wrongansatz}
 f(z) = \sqrt{z_0^2 - z^2}\,,
\end{equation}
constrained to a causal boundary condition~\cite{Hosseini:2015vya}. However, if we insert the ansatz~(\ref{eq:wrongansatz}) in this non-linear equation of motion~(\ref{eq:diffLif}), we obtain
\begin{equation} \label{eq:nogeod}
 -\frac{4 z_0^6 (\nu -1) \nu  z^3}{(z_0^2 - z^2)^{9/2}} = 0\,.
\end{equation}
It is clear that for a generic non-zero~$\nu$,~(\ref{eq:wrongansatz}) is not a solution of~ (\ref{eq:diffLif}) except the case~$\nu=1$. For~$\nu =1$, we recover the~$AdS_3$ spacetime as a special limit of the Lifshitz spacetime (in the appendix~\ref{app:Lif} we reproduce the same result using the notation of~\cite{Hosseini:2015gua}).

Since obtaining an exact solution for non-linear fourth order differential equation like~(\ref{eq:diffLif}) is generically difficult, we rather aim to solve the equation using a perturbative technique. We already know that for~$\nu =1$ we have an exact solution~(\ref{eq:wrongansatz}) of the differential equation~(\ref{eq:diffLif}).  We consider~$\nu = 1 + \d$, where~$\delta$ is a tiny positive deformation around unity and by following~\cite{Alishahiha:2013dca}\footnote{Another application of this method can be found in~\cite{Alishahiha:2013zta}} we introduce the ansatz function
\begin{equation}
 \nu = 1 + \d\,, \qquad \qquad f'(z) = h'(z) \Big(1 + \d g(z) + \d^2 n(z) + \mathcal{O}(\d^3) \Big)\,,
\end{equation}
where~$h(z) = \sqrt{z_t^2 - z^2}$ is the geodesic and~$z_t$ is the turning point. We choose the solution~(\ref{eq:geo}) for the~$AdS$ spacetime because, by taking the limit from Lifshitz to~$AdS$ (i.e by setting~$\d=0$), we end up in a region of the parameter space where the second solution~(\ref{eq:sol2}) proposed in the section~\ref{section:AdS} is not well defined ($F= 2/3$ in the language of section~\ref{section:AdS}). The downside of our choice is that the~$AdS$ central charge and the extrinsic curvature (at~$0$th-order in~$\d$) are vanishing. After imposing our ansatz, we also notice that~$g(z)$ starts to appear in the expansion of the differential equation at second order, while~$n(z)$ starts to appear at third order, and so on. Consequently, by expanding the entropy functional, the first non-trivial contribution to the entanglement entropy is coming at second order.

This ansatz will simplify the differential equation whose solution is the desired entangling surface. Following~\cite{Alishahiha:2013dca}, we can easily determine the surface by imposing few boundary conditions. In particular, we require~$z_0$ to be finite and real. Moreover, we require that our surface is anchored to the region of our interest. To do so, the condition we employ is ~$z_0 = \int_0^{z_t} f'(z)$. Finally, we also require that turning point approaches to~$0$ when the size of the entangling region~$z_0 \rightarrow 0$.

\begin{figure}[!ht]
\begin{center}
\includegraphics[scale=1]{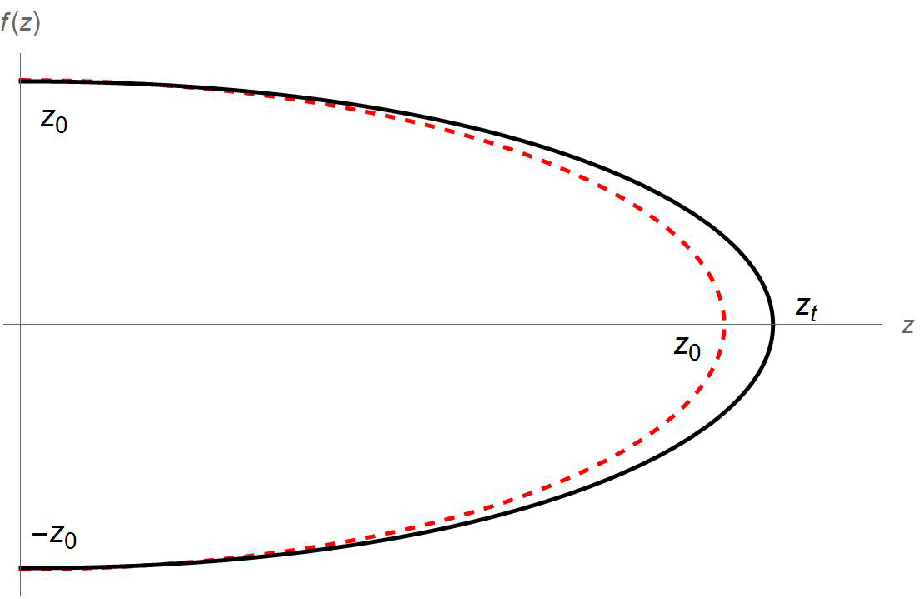}
\caption{The entangling surface (in black) is going deeper in the bulk with respect to the geodesic (in dashed red)}\label{fig:lif}
\end{center}
\end{figure}

Thanks to this approach, the differential equation becomes linear and it is solved by
\begin{equation}
 f'(z) = - \frac{z}{\sqrt{z_t^2 - z^2}} \left[ 1 + \d \left(\frac{z_t^2 ( 1 - 2 \log z_t)}{2(z_t^2 - z^2)} + \frac{z_t^2 (2 \log z - 1)}{2(z_t^2 - z^2)} \right) + \d^2 n(z) + \mathcal{O}(\d^3)  \right]\,,
\end{equation}
where the contribution at the second order is given by,
\begin{eqnarray} \label{eq:secordLif}
n(z) = &\frac{z_t^2 \left(\left(z_t^2-z^2\right)^2+z^2 \left(\left(z_t^2+2 z^2\right) (\log (z_t)-\log (z))-z_t^2+z^2\right) (\log (z_t)-\log (z))\right)}{2 \left(z^3-z\,z_t^2 \right)^2}\,.&
\end{eqnarray}
However, since~(\ref{eq:secordLif}) is not contributing at the leading order, we will not include it in the present analysis. By integrating our result, we can obtain the form of the entangling surface,
\begin{equation}
 f(z) = \sqrt{z_t^2 - z^2} + 2 z_t \d \left[ \log z - \frac{z_t \, \log (z/z_t)}{\sqrt{z_t^2 - z^2}} - \log \left(z_t + \sqrt{z_t^2 - z^2}\right) \right]  + \mathcal{O}(\d^2)\,.
\end{equation}
Now we just need to determine our turning point~$z_t$ as a function of~$z_0$, i.e. the size of the entangling region of our interest. By requiring that~$z_0 = \int_0^{z_t} f'(z)$, we obtain
\begin{equation}
 z_t = z_0 (1 + 2 \d \log 2\,)  + \mathcal{O}(\d^2)\,.
\end{equation}
It is then clear that the turning point~$z_t$ is located deeper in the bulk with respect to the one reached by the geodesic (see fig.~\ref{fig:lif}). Notice also that, as it is expected, if we take the limit~$z_0 \rightarrow 0$, the turning point~$z_t \rightarrow 0$.

Substituting these results back into the functional~(\ref{eq:eeLif}), we obtain the universal term of EE for the Lifshitz spacetime
\begin{equation}
 S_{EE}^{Lif} = \frac{c_{Lif}}{3} \log \left(\frac{z_0}{\epsilon} \right)\,, \qquad  \frac{c_{Lif}}{3} = \left[0_{AdS} - \d^2 \sqrt{\frac{3}{2}} \frac{L}{2G} + \mathcal{O}(\d^3) \right]\,.
\end{equation}
Since our expansion takes place around the chiral point of New Massive Gravity, the leading contribution (indicated as~$0_{AdS}$ to keep track of it) is vanishing and the equations are extremely simplified. However, we can see how the entangling surface is necessarily deformed around the geodesic in order to extremize the entropy functional.

From the boundary field theory point of view, the difficulties in computing the central charge for the asymptotic Lifshitz symmetry and arbitrary Lifshitz exponent have been addressed in~\cite{Compere:2009qm}. The technical reason behind this obstacle is the appearance of infinities while integrating over the non-compact~$x$ direction to obtain the conserved charge of the symmetry algebra. Therefore, it is beyond the scope of our present analysis to check our holographic results for Lifshitz spacetime from the side of boundary theory. However, from our holographic analysis, we are able to propose an approximate result of central charge as a coefficient of the leading UV logarithmic divergent term in the HEE. 

Interestingly, although the entangling surface reported in~\cite{Hosseini:2015gua} does not extremize the entropy functional (see appendix~\ref{app:Lif}), the expansion around $\nu=1$ of their result matches the one here derived. The reason is that in proximity of the boundary (where the main contribution to EE is coming from) the two entangling surfaces do not present relevant differences, as already commented in~\cite{Bueno:2014oua}.

Our analysis requires the existence of an exact solution at the zeroth order in~$\delta$, that is only known for~$\nu=1$, i.e. the chiral point. In order to explore a less simplified case, we now turn our attention to the Warped~$AdS$ spacetime as a solution of New Massive Gravity.

\section{HEE for Warped AdS$_3$ Spacetime in NMG} \label{section:WAdS}

We dedicate this section to study the effect of higher derivative contributions on holographic entanglement entropy by exploring WAdS geometry. Such spacetime has received great attention in the context of a non-AdS extension of holography that allows squashed and stretched deformations of the AdS geometry as a dual gravity spacetime. In the present discussion we are particularly interested in the timelike Warped~$AdS_3$ background having an asymptotic symmetry as~$SL(2,R) \times U(1)$. The boundary theory of such background has been proposed to be a warped conformal field theory~\cite{Detournay:2012pc,Hofman:2014loa} describing a particular class of non-Lorentzian physical systems.

The metric of the timelike WAdS$_3$ can be characterized by the $3$-dimensional version of the G\"{o}del spacetime
\begin{equation}
ds^2 = - dt^2 - 4\o r dt d\phi + 2 ( r (\ell^{-2} - \o^2)r^2) d\phi^2 + \frac{dr^2}{2 ( r (\ell^{-2} + \o^2)r^2)}\,,
\end{equation}
that solves the NMG equations of motion, provided that
\begin{equation}
m^2 \ell^2 = - \frac{19 \o^2 \ell^2 - 2}{2}\,, \qquad \qquad  \frac{\ell^2}{L^2} = \frac{11 \o^4 \ell^4 + 28 \o^2 \ell^2 - 4}{2 (19 \o^2 \ell^2 - 2)}\,.
\end{equation}
In the particular case~$\o^2\ell^2 = 1$, we find again the~$AdS_3$ spacetime. In order to simplify the discussion, following~\cite{Donnay:2015joa} and setting~$\ell=1$, we perform the necessary change of coordinates as
\begin{equation}
\theta = t - \phi\,, \qquad \qquad \rho^2 = 2r\,.
\end{equation}
With this change of coordinates, the metric takes the following form
\begin{eqnarray}
ds^2 =&& -\left(1 + (2\o - 1) \r^2 - (1 - \o^2)\frac{\r^4}{2} \right) dt^2 + 2 \left( ( \o - 1) \r^2
- (1 - \o^2) \frac{\r^4}{2} \right) dt d\theta \nn \\
&&+ \left(  \r^2
+ (1 - \o^2) \frac{\r^4}{2} \right) d \theta^2 + \frac{d\r^2}{1
	+ (1 + \o^2) \frac{\r^2}{2}}\,.
\end{eqnarray}
We choose again the boundary parametrization to describe the entangling surface
\begin{equation}
t=0\,, \qquad \qquad \theta = f(\r)\,.
\end{equation}

The curious reader will find the details of the calculation in the appendix~\ref{app:WAdS}. Intuitively, one can imagine that the complexity of the equations is forcing us to look again only for approximate solutions. It is important to notice that our choice of coordinates is made in order to recover the~$AdS$ results in the limit~$\o\rightarrow1$ at any given step of the calculation. Here we consider a particular ansatz signifying a deformation of the entangling surface for pure~$AdS_3$ spacetime in the global coordinate system, namely
\begin{equation}
\o = 1 + \d\,, \qquad \qquad f(\r) = h(\r) + \d g(\r) + \d^2 n(\r) + \mathcal{O}(\d^3)\,,
\end{equation}
where~$\d$ is a positive small deformation and
\begin{equation}
h(\r) = \tan ^{-1}\left(\frac{\sqrt{\r^2-\r_t^2}}{\r_t \sqrt{\r^2+1}}\right)\,,
\end{equation}
is the geodesic in the~$AdS_3$ background (in global coordinates), i.e the entangling surface if we set~$\o = 1$. Also in this case, we choose the geodesic of the~$AdS_3$ spacetime since, by taking~$\o=1$, we fall down in a region of the NMG parameter space where the second solution proposed in~\cite{Ghodsi:2015gna} is not well-defined.

Unlike the previous example, the function~$g(\r)$ is contributing already at the linear order in~$\d$ to the Entanglement Entropy and we are going to ignore higher order contributions. However, in order to determine the profile of~$g(\r)$, we need to solve the differential equation at order~$\d^2$. The reason is again that our ansatz, with the~$AdS_3$ geodesic at the leading order, is simplifying a lot the differential equation, forcing us to expand up to the second order to find the equation that constraints the profile~$g(\r)$.

By imposing the same boundary conditions of the previous two sections, we obtain 
\begin{eqnarray}
g(\r) &=& \frac{1}{2} \Bigg(\frac{\left(\r^2+1\right) \left(\frac{\left(\r_t^2+1\right)^3}{\r_t}-\r_t \left(4 \r_t^2+5\right) \r_{\infty}^2\right)}{\left(\r_t^2+1\right)^2 \r_{\infty}^2} \nn  \\
&&  \qquad +\frac{\r_t \left(4 \r_t^2+5\right) \r_{\infty}^2+2 \r_t \left(\r_t^2+1\right) \r_{\infty}^2 \log
	\left(\r_t^2+1\right)-\frac{\left(\r_t^2+1\right)^3}{\r_t}}{\left(\r_t^2+1\right) \r_{\infty}^2}  \nn \\
&&  \qquad +\frac{\r_t \sqrt{\r^2-\r_t^2} \left(\r_t^2 \left(4 \r^2+3\right)+5 \r^2+4\right)}{\left(\r_t^2+1\right)^2 \sqrt{\r^2+1}} \nn \\
&& \qquad  -4 \r_t \log
\left(\sqrt{\r^2-\r_t^2}+\sqrt{\r^2+1}\right)\Bigg)\,,
\end{eqnarray} 
where the turning point~$\r_t$ is determined by solving the transcendental equation
\begin{equation}
\frac{\Delta \theta}{2} = \tan^{-1} \frac{1}{\r_t} - 2 \d \r_t \log \r_{\infty}\,.
\end{equation}
Here~$\Delta \theta$ is the size of the entangling region of our interest at the UV cut-off~$\r_{\infty}$ (in other words, it is the analogue of the~$z_0$ of previous sections).

Now that we have all the ingredients, we can finally write down the universal contribution of the warping parameter to the entanglement entropy,
\begin{equation}
S_{EE}^{W} = \frac{c_{W}}{3} \log \left(\frac{\r_{\infty}}{\Delta \theta} \right)\,, \qquad\frac{c_W}{3} = \frac{4}{17 G} + \d\, \frac{52}{289 G}  + \mathcal{O}(\d^2)\,.
\label{wadscentral}
\end{equation}
In \cite{Donnay:2015iia}, by exploring the asymptotic analysis of the $WAdS_3$ spacetime in NMG, the authors have conjectured a dual WCFT$_2$. It is interesting to notice that, by expanding their result around $\omega = 1$, we can consistently reproduce the central charge we obtain in~(\ref{wadscentral}).

In this case, the leading term (the order~$\d^0$ in the expansion of the central charge, corresponding to the~$AdS$ spacetime) is non-vanishing and we can appreciate the contribution coming from the warping parameter already at first order. The nature of the deformation of the entangling surface is different from the one showed in section~\ref{section:Lif}. Although in both cases we are studying a small deformation around the exact solution of the~$AdS$ spacetime, in this case the leading order is not coming from the geodesic of the spacetime under examination, therefore there is no reason in compare the turning points of the two solutions.

\section{Discussion and Conclusions} \label{section:disclusion}

In this paper, we showed the deformation in the geometry of the entangling surface due to the presence of higher derivatives in the gravity theory. Within the perturbative approximation, we proved \emph{the existence of new entangling surfaces} for the Lifshitz and the timelike~WAdS$_3$ backgrounds in the NMG theory. In particular, the main purpose of this holographic study in the Lifshitz background is to show that, unlike~$AdS_3$ case, the Lifshitz geodesic is not the correct entangling surface that extremizes the entropy functional. Moreover, we apply the similar holographic technique in the case of timelike~WAdS$_3$ spacetime and we find an entangling surface that extremizes the entropy functional. Consequently, we compute the leading logarithmic term of holographic entanglement entropy and show that our result is consistent with the expectation of boundary Warped CFT$_2$. In both analysis, we construct new entangling surfaces as perturbative deformations over $AdS_3$ geodesic. 

One may be tempted to view the results here presented as an indication that the EE is increased or decreased as one turns on the higher derivatives. We hereby present an argument for not making such comparison in the context of our present holographic analysis.
\begin{itemize}
\item As pointed out at the end of section~\ref{section:AdS}, the correct result is the one obtained by the geodesic, further explanation of why the second result, which was giving a lower entanglement entropy, should be discharged can be found in~\cite{Mozaffar:2016hmg}. 
\item Regarding the Lifshitz case, it is not appropriate to interpret the result as a reduction in the EE, since the 0th order, i.e. the geodesic (both of the Lifshitz and the AdS spacetime) is not an admissible surface to compute the EE in Lifshitz spacetime (see equation~\ref{eq:nogeod}).
\item For the~$WAdS_3$ spacetime, we can interpret our result as a perturbative enhancement of the holographic entanglement entropy due to warping of the AdS spacetime. However, the comparison is not very meaningful since the entangling at the zeroth order does not represent any particular surface for the WAdS$_3$ spacetime (being simply the AdS geodesic).
\end{itemize}

In~\cite{Bueno:2014oua}, the authors showed how the presence of higher derivative terms in the gravity theory does not change the structure of the divergences in the entanglement entropy with respect to the Einstein gravity case. However, since the backgrounds taken into examination are not solutions of the pure Einstein gravity, we find more appropriate to show explicitly the shifts in the central charges. These changes in the shape of the entangling surface, as well as in the coefficient of the leading divergence in the entanglement entropy, are present exclusively because we take into account the higher derivative terms. To this purpose, the Lifshitz case, presented in section~\ref{section:Lif}, is a perfect example, since \emph{the surface determined by ignoring the higher derivative terms (i.e. the geodesic) is not extremizing the entropy functional}.

As suggested in~\cite{Erdmenger:2014tba}, the technical reason behind the emergence of multiple entangling surfaces is that we can't impose a sufficient number of boundary conditions to solve our differential equations. In~\cite{Hosseini:2015vya}, the authors proposed the so-called \emph{free-kick condition} in order to solve the problem in the context of hairy black holes as solutions of NMG. However, such condition constrains the entangling surface to be the geodesic at its turning point and we can find counter examples both in~\cite{Ghodsi:2015gna} and in the analysis here presented. 

We believe that the central question to be addressed in the near future is how can we give a physical reason that solves such a technical problem. We provide explicit (although approximate) solutions and we thereby hope that this work will pave the way for a discussion to find a rigorous method to solve such problems.
 
In~\cite{Song:2016pwx}, the computation of the holographic entanglement entropy for WAdS spacetime has been investigated in the light of non-AdS holography. It would be very interesting to check if the non-AdS correction considered in~\cite{Song:2016pwx} can be consistently implemented together with higher derivative contribution of the NMG theory. Moreover, same kind of analysis can be performed for geometries with a horizon (i.e. black holes). We reserve these problems for our future study.

\section*{Acknowledgements}

We thank Eric Bergshoeff for his guidance and supervision during the entire process.  We also thank Gokhan Alkac for the useful discussions and the help in some of the calculations. It is a pleasure to acknowledge Arjun Bagchi, Souvik Banerjee, Rudranil Basu and Wei Song for various fruitful discussions and valuable comments on the draft. The research of LB is supported by the Dutch stichting voor Fundamenteel Onderzoek der Materie (FOM). SC is supported by Erasmus Mundus NAMASTE India-EU Grants during a major part of this work.

\appendix

\section{Details of the Lifshitz case} \label{app:Lif}

We dedicate this appendix to review the calculation presented in~\cite{Hosseini:2015gua}, using their notation to avoid confusion. The metric is given by
\begin{equation}
d s^2 = g_{\m\n} d x^{\m} dx^{\n} = -\frac{r^{2\nu}}{\widetilde{L}^{2\nu}} dt^2 + \frac{\widetilde{L}^{2}}{r^{2}} dr^2 + r^2 d\phi^2\,,
\end{equation}
where
\begin{equation}
m^2 \widetilde{L}^2 = \frac{1}{2} \left( \nu^2 - 3 \nu + 1 \right)\,, \qquad \qquad \frac{\widetilde{L}^2}{L^2} = \frac{1}{2} \left( \nu^2 + \nu + 1 \right)\,.
\end{equation}
The induced metric, the Ricci scalar and~$R_{||}$ are given by
\begin{equation}
h = \frac{\widetilde{L}^2}{r^2} + r^2 f'(r)^2 \,, \quad  R = -\frac{2\left( \nu^2 + \nu + 1 \right)}{\widetilde{L}^2}\,, \quad  R_{||} = \frac{(\nu-1) \nu}{\widetilde{L}^2 + r^4 f'(r)^2} - \frac{2\nu^2 + \nu + 1}{\widetilde{L}^2}\,,
\end{equation}
and the last term we need is
\begin{equation}
K^2 = \frac{r^4 \left(\widetilde{L}^2 \big(r f''(r)+3 f'(r)\big)+r^4 f'(r)^3\right)^2}{\widetilde{L}^2 \left(\widetilde{L}^2+r^4 f'(r)^2\right)^3}\,.
\end{equation}
In this notation, the differential equation to be solved is
\begin{eqnarray}
&&r^{12} \widetilde{L}^2 f'(r)^6 \left(\left(-4 \nu ^2+6 \nu -3\right) r f''(r)-3 (2 (\nu -3) \nu +3) f'(r)\right)  \nn\\
&&-r^8 \widetilde{L}^4 f'(r)^2 \left(30 r^3
f''(r)^3+(6 (\nu -5) \nu +99) f'(r)^3+5 r^2 f'(r) f''(r) \left(15 f''(r)-4 r f^{(3)}(r)\right) \right. \nn \\
&& \left. \qquad \qquad \qquad +2 r f'(r)^2 \left(3 ((\nu -2) \nu +19) f''(r)+r
\left(r f^{(4)}(r)-10 f^{(3)}(r)\right)\right)\right) \nn \\
&&+r^4 \widetilde{L}^6 \left(5 r^3 f''(r)^3+(2 \nu  (3 \nu +7)+165) f'(r)^3+r f'(r)^2 \left(3 (2
\nu +87) f''(r)-4 r^2 f^{(4)}(r)\right) \right. \nn \\
&& \left. \qquad \qquad +5 r^2 f'(r) f''(r) \left(4 r f^{(3)}(r)+27 f''(r)\right)\right) \nn \\
&&+2 \widetilde{L}^8 \left(3 \left(\nu
^2-4\right) f'(r)-r \left(\left(24-\nu ^2\right) f''(r)+r \left(r f^{(4)}(r)+10 f^{(3)}(r)\right)\right)\right) \nn \\
&&+(2 \nu -1) r^{16} f'(r)^9 = 0\,.
\end{eqnarray}
If we take the entangling surface to be the geodesic
\begin{equation}
f(r) = \frac{\widetilde{L} \sqrt{r^2 \widetilde{L}^2-r_t^2}}{r\, r_t}\,,
\end{equation}
we obtain
\begin{equation}
\frac{4\, \widetilde{L}^{15} r_t^3\, (\nu -1)\, \nu \, r^4 }{\left(\widetilde{L}^2 r^2 -r_t^2\right)^{9/2}} = 0\,.
\end{equation}
Therefore we can conclude that the geodesic cannot be taken as entangling surface in a Lifshitz background, since it doesn't minimize the entropy functional, except for the case~$\nu=1$, which corresponds to the Anti-de Sitter spacetime.

\section{Details of the WAdS case} \label{app:WAdS}

In this appendix, we present the details of our calculation of section~\ref{section:WAdS}. We choose the boundary parametrization of the entangling surface, i.e.~$t=0$ and~$\theta = f(\r)$. Thus the induced metric is given by
\begin{equation}
h = \r^2 \left(\frac{1}{2} \r^2 \left(\omega ^2-1\right)+1\right) f'(\r)^2+\frac{1}{\frac{1}{2} \r^2 \left(\omega ^2+1\right)+1}\,.
\end{equation}
We compute the orthogonal vectors
\begin{eqnarray}
 n_{\a\,1} &=& C \left(-\frac{(\omega -1) \left(\r^2 (\omega +1)+2\right)}{\r^2 \left(\omega ^2-1\right)-2},-f'(\r),1\right)\,, \nn \\
 n_{\a\,2} &=& \left(-\sqrt{\left| \frac{\left(\omega ^2+1\right) \r^2+2}{2-\r^2 \left(\omega ^2-1\right)}\right| },0,0\right)\,, \nn \\
 C &=& \frac{\sqrt{2}\, \r}{\sqrt{\r^2 \left(\r^2 \left(\omega ^2+1\right)+2\right) f'(\r)^2+\frac{4}{2-\r^2 \left(\omega ^2-1\right)}}}\,,
\end{eqnarray}
as well as the contributions of the intrinsic curvature (remember that we set~$\ell=1$)
\begin{eqnarray}
&R = -2 \left(\omega ^2+2\right)\,,  & \\
&R_{||} = \frac{2 \r^2 \left(\r^2 \left(\omega ^2-1\right)^2+2 \left(\omega ^2+1\right)\right) \left(\r^2 \left(\omega ^2+1\right)+2\right)
	f'(\r)^2+8 \left(\omega ^2+1\right)}{\r^2 \left(\r^4 \omega ^4-\left(\r^2+2\right)^2\right) f'(\r)^2-4}\,,
\end{eqnarray}
and the extrinsic curvature contraction 
\begin{eqnarray}
K^2 &=&\frac{2}{\left(\r^2 \left(\r^4 \omega ^4-\left(\r^2+2\right)^2\right) f'(\r)^2-4\right)^3}  \left[4 \r^2 \left(\r^2 \left(\omega ^2-1\right)-2\right) \left(\r^2 \left(\omega ^2+1\right)+2\right)^2 f''(\r)^2 \right. \nn \\
&& \left. +\r^4 \left(\r^2 \left(\omega
^2-1\right)-2\right) \left(\r^2 \left(\omega ^2-1\right)-1\right)^2 \left(\r^2 \left(\omega ^2+1\right)+2\right)^4 f'(\r)^6 \right. \nn \\
&& \left.-4 \r^2
\left(\r^2 \left(\omega ^2-1\right)-2\right) \left(\r^2 \left(\omega ^2+1\right)+2\right)^2 \left(\r^4 \left(3 \omega ^4+2 \omega
^2-5\right) \right. \right. \nn\\
&&\left. \left.+3 r^2 \left(\omega ^2-3\right)-4\right) f'(\r)^4 +4 \left(\r^6  \left(\omega ^4-1\right) \left(17
\omega ^2+25\right)  \right. \right. \nn \\
&&\left. \left. +\r^4 \left(38 \omega ^4-20 \omega ^2-90\right)-96 \r^2-32\right) f'(\r)^2 \right. \nn \\
&& \left. +8 \r \left(\r^2 \left(\omega
^2+1\right)+2\right) \left(5 \r^4 \left(\omega ^4-1\right)+2 \r^2 \left(\omega ^2-7\right)-8\right) f'(\r) f''(\r) \right. \nn \\
&&\left. -4 \r^3 \left(\r^2
\left(\omega ^2-1\right)-2\right) \left(\r^2 \left(\omega ^2-1\right)-1\right) \left(\r^2 \left(\omega ^2+1\right)+2\right)^3 f'(\r)^3
f''(\r) \right]\,.
\end{eqnarray}
The merit of our choice of coordinates is that all these quantities, in the limit~$\o\rightarrow1$, are precisely and smoothly the one that one obtain for the~$AdS_3$ case (in global coordinates). With these results we can write down the entropy functional~(\ref{eq:EEfunc})
\begin{eqnarray}
S_{EE} &=& \frac{1}{4 G} \int d\r\, \sqrt{h}\, \Bigg(1-\frac{2}{19 \omega ^2-2}\, \Bigg(\frac{3}{2} \left(\omega ^2+2\right) + \frac{1}{\left(\r^2 \left(\r^4 \omega ^4-\left(\r^2+2\right)^2\right) f'(\r)^2-4\right)^3}  \nn \\
&& \qquad \qquad \left(-4 \r^2 \left(\r^2 \left(\omega ^2-1\right)-2\right) \left(\r^2 \left(\omega ^2+1\right)+2\right)^2 f''(\r)^2 \right. \nn \\
&& \qquad \qquad +\r^4 \left(\r^2 \left(\omega
^2-1\right)-2\right) \left(\r^2 \left(\omega ^2+1\right)+2\right)^3 \left(\r^2 \left(\r^2 \left(\omega ^2-1\right) \right.\right. \nn \\
&& \qquad \qquad \left.\left.\left(\r^2
\left(\omega ^4-4 \omega ^2+3\right)+12\right)-5 \omega ^2-13\right)-2\right) f'(\r)^6 \nn \\
&& \qquad \qquad +4 \r^2 \left(\r^2 \left(\omega
^2-1\right)-2\right) \left(\r^2 \left(\omega ^2+1\right)+2\right)^2 \nn \\
&& \qquad \qquad \left(\r^4 \left(\omega ^4+10 \omega ^2-11\right)-3 \r^2 \left(3
\omega ^2+7\right)-4\right) f'(\r)^4 \nn \\
&& \qquad \qquad  -4 \left(\r^6 \left(\omega^4 -1\right) \left(25 \omega ^2+49\right)+6 \r^4
\left(\omega ^4-14 \omega ^2-31\right) -96 \r^2 \left(\omega ^2+2\right)-32\right) f'(\r)^2 \nn\\
&& \qquad \qquad -8 \r \left(\r^2 \left(\omega
^2+1\right)+2\right) \left(5 \r^4 \left(\omega ^4-1\right)+2 \r^2 \left(\omega ^2-7\right)-8\right) f'(\r) f''(\r) \nn \\
&& \qquad \qquad +4 \r^3 \left(\r^2
\left(\omega ^2-1\right)-2\right) \left(\r^2 \left(\omega ^2-1\right)-1\right) \left(\r^2 \left(\omega ^2+1\right)+2\right)^3 f'(\r)^3
f''(\r) \nn \\
&& \qquad \qquad +128 \left(\omega ^2+1\right) \Big) \Bigg) \Bigg)\,.
\end{eqnarray}
To extremize the action, we need to solve the equation of motion derived from this functional. However, as in the Lifshitz case, the resulting differential equation is highly non-linear. Therefore, we solve such complex equation with the perturbative techniques of section~\ref{section:WAdS}.




\begin{thebibliography}{99}

\bibitem{Calabrese:2004eu}
  P.~Calabrese and J.~L.~Cardy,
  ``Entanglement entropy and quantum field theory,''
  J.\ Stat.\ Mech.\  {\bf 0406} (2004) P06002
  [hep-th/0405152].
  
   \bibitem{Casini:2009sr} 
   H.~Casini and M.~Huerta,
   ``Entanglement entropy in free quantum field theory,''
   J.\ Phys.\ A {\bf 42}, 504007 (2009)
   [arXiv:0905.2562 [hep-th]].
   
   \bibitem{Ryu:2006bv} 
   S.~Ryu and T.~Takayanagi,
   ``Holographic derivation of entanglement entropy from AdS/CFT,''
   Phys.\ Rev.\ Lett.\  {\bf 96}, 181602 (2006)
   [hep-th/0603001].
   
   \bibitem{Ryu:2006ef} 
   S.~Ryu and T.~Takayanagi,
   ``Aspects of Holographic Entanglement Entropy,''
   JHEP {\bf 0608}, 045 (2006)
   [hep-th/0605073].
  
  \bibitem{Nishioka:2009un}
  T.~Nishioka, S.~Ryu and T.~Takayanagi,
  ``Holographic Entanglement Entropy: An Overview,''
  J.\ Phys.\ A {\bf 42} (2009) 504008
  [arXiv:0905.0932 [hep-th]].
  
  \bibitem{Hubeny:2007xt} 
  V.~E.~Hubeny, M.~Rangamani and T.~Takayanagi,
  ``A Covariant holographic entanglement entropy proposal,''
  JHEP {\bf 0707}, 062 (2007)
  [arXiv:0705.0016 [hep-th]].  
  
  \bibitem{VanRaamsdonk:2010pw}
  M.~Van Raamsdonk,
  ``Building up spacetime with quantum entanglement,''
  Gen.\ Rel.\ Grav.\  {\bf 42} (2010) 2323
  [Int.\ J.\ Mod.\ Phys.\ D {\bf 19} (2010) 2429]
  [arXiv:1005.3035 [hep-th]].
  
  \bibitem{Swingle:2009bg}
  B.~Swingle,
  ``Entanglement Renormalization and Holography,''
  Phys.\ Rev.\ D {\bf 86} (2012) 065007
  [arXiv:0905.1317 [cond-mat.str-el]].
  
  \bibitem{Lewkowycz:2013nqa}
  A.~Lewkowycz and J.~Maldacena,
  ``Generalized gravitational entropy,''
  JHEP {\bf 1308} (2013) 090
  [arXiv:1304.4926 [hep-th]].
  
  \bibitem{Dong:2013qoa} 
  X.~Dong,
  ``Holographic Entanglement Entropy for General Higher Derivative Gravity,''
  JHEP {\bf 1401}, 044 (2014)
  [arXiv:1310.5713 [hep-th]].
  
  \bibitem{Fursaev:2013fta} 
  D.~V.~Fursaev, A.~Patrushev and S.~N.~Solodukhin,
  ``Distributional Geometry of Squashed Cones,''
  Phys.\ Rev.\ D {\bf 88}, no. 4, 044054 (2013)
  [arXiv:1306.4000 [hep-th]].
  
  \bibitem{Camps:2013zua} 
  J.~Camps,
  ``Generalized entropy and higher derivative Gravity,''
  JHEP {\bf 1403}, 070 (2014)
  [arXiv:1310.6659 [hep-th]].  
  
  \bibitem{Wald:1993nt}
  R.~M.~Wald,
  ``Black hole entropy is the Noether charge,''
  Phys.\ Rev.\ D {\bf 48} (1993) 3427
  [gr-qc/9307038].
  
  \bibitem{Bergshoeff:2009hq}
  E.~A.~Bergshoeff, O.~Hohm and P.~K.~Townsend,
  ``Massive Gravity in Three Dimensions,''
  Phys.\ Rev.\ Lett.\  {\bf 102} (2009) 201301
  [arXiv:0901.1766 [hep-th]].
  
  \bibitem{Erdmenger:2014tba}
  J.~Erdmenger, M.~Flory and C.~Sleight,
  ``Conditions on holographic entangling surfaces in higher curvature gravity,''
  JHEP {\bf 1406} (2014) 104
  [arXiv:1401.5075 [hep-th]].

  \bibitem{Ghodsi:2015gna}
  A.~Ghodsi and M.~Moghadassi,
  ``Holographic entanglement entropy from minimal surfaces with/without extrinsic curvature,''
  JHEP {\bf 1602} (2016) 037
  [arXiv:1508.02527 [hep-th]].
  
  \bibitem{Alishahiha:2013dca} 
  M.~Alishahiha, A.~F.~Astaneh and M.~R.~M.~Mozaffar,
  ``Holographic Entanglement Entropy for 4D Conformal Gravity,''
  JHEP {\bf 1402}, 008 (2014)
  [arXiv:1311.4329 [hep-th]].  

  \bibitem{Bhattacharyya:2013gra}
  A.~Bhattacharyya, M.~Sharma and A.~Sinha,
  ``On generalized gravitational entropy, squashed cones and holography,''
  JHEP {\bf 1401} (2014) 021
  [arXiv:1308.5748 [hep-th]].

  \bibitem{Hosseini:2015vya}
  S.~M.~Hosseini and A.~Veliz-Osorio,
  ``Free-kick condition for entanglement entropy in higher curvature gravity,''
  Phys.\ Rev.\ D {\bf 92} (2015) 4,  046010
  [arXiv:1505.00826 [hep-th]].
  
  \bibitem{Hosseini:2015gua} 
  S.~M.~Hosseini and A.~Veliz-Osorio,
  ``Entanglement and mutual information in 2d nonrelativistic field theories,''
  Phys.\ Rev.\ D {\bf 93} (2016) no.2,  026010
  [arXiv:1510.03876 [hep-th]].
  
  \bibitem{Anninos:2013nja}
  D.~Anninos, J.~Samani and E.~Shaghoulian,
  ``Warped Entanglement Entropy,''
  JHEP {\bf 1402} (2014) 118
  [arXiv:1309.2579 [hep-th]].
  
  \bibitem{Castro:2015csg}
  A.~Castro, D.~M.~Hofman and N.~Iqbal,
  ``Entanglement Entropy in Warped Conformal Field Theories,''
  JHEP {\bf 1602} (2016) 033
  [arXiv:1511.00707 [hep-th]].
  
  \bibitem{Ghodrati:2016ggy}
  M.~Ghodrati and A.~Naseh,
  ``Phase transitions in BHT Massive Gravity,''
  arXiv:1601.04403 [hep-th].

  \bibitem{Fischler:2012ca} 
  W.~Fischler and S.~Kundu,
  ``Strongly Coupled Gauge Theories: High and Low Temperature Behavior of Non-local Observables,''
  JHEP {\bf 1305}, 098 (2013)
  [arXiv:1212.2643 [hep-th]].
  
  \bibitem{Mozaffar:2016hmg}
  M.~R.~M.~Mozaffar, A.~Mollabashi, M.~M.~Sheikh-Jabbari and M.~H.~Vahidinia,
  ``Holographic Entanglement Entropy, Field Redefinition Invariance and Higher Derivative Gravity Theories,''
  arXiv:1603.05713 [hep-th].

  \bibitem{Hartong:2014oma}
  J.~Hartong, E.~Kiritsis and N.~A.~Obers,
  ``Lifshitz space–times for Schrödinger holography,''
  Phys.\ Lett.\ B {\bf 746} (2015) 318
  [arXiv:1409.1519 [hep-th]].
  
  \bibitem{Griffin:2012qx}
  T.~Griffin, P.~Hořava and C.~M.~Melby-Thompson,
  ``Lifshitz Gravity for Lifshitz Holography,''
  Phys.\ Rev.\ Lett.\  {\bf 110} (2013) no.8,  081602
  [arXiv:1211.4872 [hep-th]].
  
  \bibitem{Son:2008ye}
  D.~T.~Son,
  ``Toward an AdS/cold atoms correspondence: A Geometric realization of the Schrodinger symmetry,''
  Phys.\ Rev.\ D {\bf 78} (2008) 046003
  [arXiv:0804.3972 [hep-th]].
  
  \bibitem{Balasubramanian:2008dm}
  K.~Balasubramanian and J.~McGreevy,
  ``Gravity duals for non-relativistic CFTs,''
  Phys.\ Rev.\ Lett.\  {\bf 101} (2008) 061601
  [arXiv:0804.4053 [hep-th]].
  
  \bibitem{Kachru:2008yh}
  S.~Kachru, X.~Liu and M.~Mulligan,
  ``Gravity duals of Lifshitz-like fixed points,''
  Phys.\ Rev.\ D {\bf 78} (2008) 106005
  [arXiv:0808.1725 [hep-th]].

  \bibitem{Korovin:2013bua}
  Y.~Korovin, K.~Skenderis and M.~Taylor,
  ``Lifshitz as a deformation of Anti-de Sitter,''
  JHEP {\bf 1308} (2013) 026
  [arXiv:1304.7776 [hep-th]].

  \bibitem{deBoer:2011wk}
  J.~de Boer, M.~Kulaxizi and A.~Parnachev,
  ``Holographic Entanglement Entropy in Lovelock Gravities,''
  JHEP {\bf 1107} (2011) 109
  [arXiv:1101.5781 [hep-th]].
  
  \bibitem{AyonBeato:2009nh}
  E.~Ayon-Beato, A.~Garbarz, G.~Giribet and M.~Hassaine,
  ``Lifshitz Black Hole in Three Dimensions,''
  Phys.\ Rev.\ D {\bf 80} (2009) 104029
  [arXiv:0909.1347 [hep-th]].
  
  \bibitem{Alishahiha:2013zta}
  M.~Alishahiha, A.~F.~Astaneh and M.~R.~M.~Mozaffar,
  ``Entanglement Entropy for Logarithmic Conformal Field Theory,''
  Phys.\ Rev.\ D {\bf 89} (2014) no.6,  065023
  [arXiv:1310.4294 [hep-th]].

  \bibitem{Compere:2009qm} 
  G.~Compere, S.~de Buyl, S.~Detournay and K.~Yoshida,
  ``Asymptotic symmetries of Schrodinger spacetimes,''
  JHEP {\bf 0910}, 032 (2009)
  [arXiv:0908.1402 [hep-th]].
  
  \bibitem{Bueno:2014oua} 
  P.~Bueno and P.~F.~Ramirez,
  ``Higher-curvature corrections to holographic entanglement entropy in geometries with hyperscaling violation,''
  JHEP {\bf 1412}, 078 (2014)
  [arXiv:1408.6380 [hep-th]].

  \bibitem{Detournay:2012pc}
  S.~Detournay, T.~Hartman and D.~M.~Hofman,
  ``Warped Conformal Field Theory,''
  Phys.\ Rev.\ D {\bf 86} (2012) 124018
  [arXiv:1210.0539 [hep-th]].
  
  \bibitem{Hofman:2014loa}
  D.~M.~Hofman and B.~Rollier,
  ``Warped Conformal Field Theory as Lower Spin Gravity,''
  Nucl.\ Phys.\ B {\bf 897} (2015) 1
  [arXiv:1411.0672 [hep-th]].

  \bibitem{Donnay:2015joa}
  L.~Donnay, J.~J.~Fernandez-Melgarejo, G.~Giribet, A.~Goya and E.~Lavia,
  ``Conserved charges in timelike warped AdS$_3$ spaces,''
  Phys.\ Rev.\ D {\bf 91} (2015) no.12,  125006
  [arXiv:1504.05212 [hep-th]].
 
  \bibitem{Donnay:2015iia}
  L.~Donnay and G.~Giribet,
  ``Holographic entropy of Warped-AdS$_{3}$ black holes,''
  JHEP {\bf 1506} (2015) 099
  [arXiv:1504.05640 [hep-th]].
  
  \bibitem{Song:2016pwx} 
  W.~Song, Q.~Wen and J.~Xu,
  ``Generalized Gravitational Entropy for WAdS3,''
  arXiv:1601.02634 [hep-th].

 \end{thebibliography}
\end{document}